# Optically detected magnetic resonance study of thermal effects due to absorbing environment around nitrogen - vacancy - nanodiamond powders

**Mona Jani,**[1,2,a] **Zuzanna Orzechowska,**[1] **Mariusz Mrózek,**[1] **Marzena Mitura-Nowak,**[3] **Wojciech Gawlik**[1] **and Adam M. Wojciechowski**[1,a]

## AFFILIATIONS

[1]Marian Smoluchowski Institute of Physics, Jagiellonian University, 30-348 Kraków, Poland

[2]Laser Centre, University of Latvia, Jelgavas street 3, LV-1004, Riga, Latvia

[3]Institute of Nuclear Physics, Polish Academy of Sciences, Radzikowskiego 152, 31-342 Kraków, Poland

[a]Author to whom correspondence should be addressed: mona.jani@lu.lv and a.wojciechowski@uj.edu.pl

## ABSTRACT

We implanted $Fe^+$ ions in nanodiamond (ND) powder containing negatively charged nitrogen-vacancy ($NV^-$) centers and studied their Raman spectra and optically detected magnetic resonance (ODMR) in various applied magnetic fields with green light (532 nm) excitation. In Raman spectra, we observed a *blue shift* of the $NV^-$ peak associated with the conversion of the electronic $sp^3$ configuration to the disordered $sp^2$ one typical for the carbon/graphite structure. In the ODMR spectra, we observed a *red shift* of the resonance position caused by local heating by an absorptive environment that recovers after annealing. To reveal the *red shift* mechanism in ODMR, we created a controlled absorptive environment around ND by adding iron-based $Fe_2O_3$ and graphitic $sp^2$ powders to the ND suspension. This admixture caused a substantial increase in the observed shift proportional to the applied laser power, corresponding to an increase in the local temperature by 150-180 K. This surprisingly large shift is absent in non-irradiated NV-ND powders, is associated only with the modification of the local temperature by the absorptive environment of NV-NDs and can be studied using ODMR signals of $NV^-$.

*Keywords:* magnetic resonance; nitrogen-vacancy; nanodiamond; thermal effects; non-carbon environment

## I. INTRODUCTION

Taking advantage of their unique quantum properties, nitrogen-vacancy (NV) color centers in nanodiamonds (NDs) have been research strategy for highly sensitive and spatially resolved room temperature nanothermometry and nanomagnetometry applications[1–4] with numerous applications, including those studied for electronic devices and biological sample objects that can operate under ambient and extreme conditions[2,5–7]. NV color centers in ND occur in two optically distinguishable electronic states: negatively charged (NV⁻) and neutral (NV⁰). When excited with green light (532 nm), the photoluminescence (PL) contains signatures of both centres, NV⁻ and NV⁰. Specifically, the sharp lines at 637 and 575 nm represent the zero-phonon lines (ZPL) of NV⁻ and NV⁰, respectively. In the PL spectrum, the relative ratio of NV⁻ to NV⁰ for each charge state depends on the excitation wavelengths, as discussed in numerous articles, e.g. in Ref. [8]. For green light (532 nm) excitation, the ZPL of NV⁻ is significantly stronger than that of NV⁰[8]. NV⁻ has a nonzero electronic spin ($S = 1$) and a triplet ground level[9]. The superior photostability of NV⁻ enables precise optical detection of the electron spin resonance and, thus, conducting fundamental studies, such as on interactions with the nearby electronic and nuclear spins. Most of the existing applications of ND with NV centers explore the exceptional optical and spin properties of NV⁻, in particular their dependence on electric[10] and magnetic fields[6], strain[11], and temperature[12]. After illumination with green light, the NV⁻ centers emit a spin-state-dependent fluorescence. The optically detected magnetic resonance (ODMR) technique relies on recording the fluorescence intensity while driving the spin transitions with appropriate microwave (MW) fields. This allows us to read out the populations of the ground-state magnetic sublevels, $m_s = 0$ and $\pm 1$ split by the zero-field splitting (ZFS) parameter $D$ (equal to about 2870 MHz at zero magnetic field) at room temperature. The dependence of the energies of the states $m_s = \pm 1$ on the external magnetic field enables many sensing applications[13], also on the nanoscale. For nanoscale temperature sensing/monitoring, one uses the fact that the ZFS parameter $D$ decreases in response to a temperature increase due to lattice thermal expansion[12], whereas for nanoscale magnetic field sensing with arbitrary oriented NDs, such as in powders or suspensions, one can benefit from a broadening of the overall ODMR spectrum that enables determination of the external magnetic-field strength[14].

NV⁻ centers have been widely used to investigate magnetic nanostructures[15], paramagnetic ions in solution[16], spin-labelled molecules, detect temperature changes[17], functionalized surfaces, measurements within living cells[18], increase the sensitivity of biosensors, etc.[19]. Most applications that use NV-ND rely on interactions between the NV⁻ centers and the environment close to the ND surface[20]. NV⁻ centers provide significant information on various environmental influences[21,22,7,17]. For example, the highly sensitive surface of ND enables complex surface immobilizations, grafting chemical moieties or biomolecules that strongly affect the surface reactivity of NDs, and can even induce chemical contaminations or surface graphitization[23,24,2,25]. The presence of non-diamond carbon impurities reduces the fluorescence efficiency of ND[26]. In addition, the fluorescence intensity of NV-NDs depends on their ambient magnetic and fluctuating environment. Defects on the surface of the ND, such as hydrogen, vacancies, and trapping states, reduce the stability of NV⁻ centers[25,27,28]. Because for many NV-ND-based applications, a very specific, uniform, and homogeneous surface is vitally important, we apply ODMR signals of NV⁻ to study the environment around the NV-ND powder surface.

In this work, we implanted NDs with Fe ions. The use of metal ions that transform to the $sp^2$-graphitic phase is reported for field emission sources[29], and as magnetic biomaterials[30], and this was the motivation to start initially with the implantation of metal - Fe ions. When performing spectral diagnostic of the Fe irradiated samples, we observed a characteristic *blue shift* of the 1570 cm⁻¹ Raman line typical for a graphitic/non-carbon phase and a large additional *red shift* of the ODMR. To obtain a closer view of the observed shifts, we admixed controlled amounts of iron-based $Fe_2O_3$ and graphitic ($sp^2$) powders into the ND suspension used for sample preparation. This admixture caused a substantial increase in the observed *red shift* in the magnetic resonance proportional to the applied laser power, corresponding to an increase in the local temperature by 150-180 K. This resonance shift is absent in non-irradiated NV-ND powders and is associated only with the modification of the local temperature by the absorptive environment of NV-NDs. Although it is generally well known that admixtures and contaminations cause enhanced absorption of external radiation (optical and MW) in the pure crystal structure, the huge size of the observed local heating studied using ODMR was quite unexpected and motivated this specific detailed study.

Our results confirm that the shift is caused by thermal effects arising from our modification of the ND environment and can be detrimental to sensitive measurements. Dried NDs are poorly thermally anchored and can heat up significantly before dissipating the heat elsewhere. While our focus was on implantation of Fe ions, we recognize that similar outcomes could be anticipated for various types of ions. We hope that this work contributes to a specific understanding of the interaction between NV⁻ spins in NDs and the surrounding environment and should be useful for implementations of thermometry and magnetometry methods and the development of NV relaxometry for material and bioscience research.

## II. EXPERIMENTAL METHOD AND SETUP

We used a 140-nm carboxylated ND slurry with a mean diamond particle size of 140 nm, having a concentration of 1.5 ppm of NV centres, dispersed in deionized water (DI) (1 mg/mL) obtained from Adamas nanotechnologies, US. The spectroscopic properties of these NDs have been previously characterized in Ref. 31. From the suspension, 40 µl was dropped onto precleaned glass substrates and left to form a dry deposit.

Subsequently, the samples were implanted with $Fe^+$ ions using the ion implanter at the Institute of Nuclear Physics (Kraków Poland) described in Ref. 32. The implantation was carried out with ions having 30 keV energy and current density in the range of 0.8 - 1.0 µA/cm$^2$, with substantial ion doses: $10^{14}$ and $10^{15}$ ions/cm$^2$. We labelled these samples as ND14 and ND15, respectively. For comparison of samples before and after the implantation process, samples were annealed in vacuum at 600 °C for 3 hours.

To create an iron-based and graphitic environment around NV-ND, we separately mixed and thoroughly sonicated ~ 3 mg of iron containing $Fe_2O_3$ (Warchem, Poland) and graphite powders in 2 ml of DI. From this 75 µl was mixed with 50 µl of ND slurry and 40 µl was drop-dried on precleaned glass substrates. We labelled them as NDF and NDG, respectively.

Here, we use five samples that are: non-irradiated NV-ND; non-irradiated NV-NDs implanted with $Fe^+$ ions at dose $10^{14}$ ions/cm$^2$ (ND14) and $10^{15}$ ions/cm$^2$ (ND15); and NDs embedded with $Fe_2O_3$ (NDF) and graphitic (NDG). ND14 and ND15 were annealed for comparison.

Depth profile simulations for $^{56}Fe$ ion implantation in NDs were obtained by a popular simulation routine Stopping and Range of Ions in Matter (SRIM)[33]. Raman scattering spectra were recorded with laser excitation of 532 nm in the range of 110 - 4000 cm$^{-1}$ with a spectral resolution of 2 cm$^{-1}$ using a confocal micro-Raman spectrometer (Almega XR, Thermo Electron Corp.) and WITec Alpha 300 confocal Raman microscope (Ulm, Germany).

ODMR signals were obtained in a home-built wide-field epifluorescence microscopy setup[34]. Samples on glass substrates were mounted close to the printed-circuit board of MW and excited by a 532 nm green laser (Sprout G, Lighthouse Photonics) with ~ 65 mW laser power focused onto the back-focal plane of the 40×/0.6 objective (LUCPLFLN, Olympus). The continuous wave MW field was generated by a signal generator (SG386, Stanford Research Systems) plus the (ZHL-16W-43+ Mini-Circuits) amplifier and its frequency was swept around the resonance frequency of the $m_s = 0 \leftrightarrow m_s = \pm 1$ transition of the ground state of $NV^-$ of about 2870 MHz with 1 MHz steps. An external uniform magnetic field ($B$) of 1 mT and 3 mT was applied to the sample using a permanent magnet. To elucidate the heating effects due to MW and laser on the samples, the ODMR signals were collected by changing the MW power between -30 and +10 dBm at a constant laser power of ~ 65 mW and by varying the laser power between 5 and 80 mW at a constant MW power of -10/-20 dBm (before amplifier). The red fluorescence (~ 600-800 nm) emitted from the sample was collected with the same microscope objective, filtered and projected onto a 704 x 594 pixel photosensitive sensor of the sCMOS camera (Andor Zyla 5.5). The objective used produced an effective pixel size of about (0.29 µm)$^2$ with a field of view of 205 µm x 173 µm. The ODMR data collection was carried out in LabView software, which controlled the camera, the signal generator, and the digital pattern generator (Pulse Streamer 8/2, Swabian Instruments). The laser spot on the sample has an area of 1.02 x 10$^5$ µm$^2$ and the laser energy flux on the NV-NDs is ~ 6.3 x 10$^{-7}$ W/µm$^2$. During continuous light illumination, two images were captured with MWs turned on and off for every frequency to subtract the background and calculate the normalized fluorescence contrast, which helped to reduce fluorescence intensity variations due to laser power drifts and improved the overall signal-to-noise ratio.

## III. RESULTS AND DISCUSSION

### A. Implantation: Raman spectroscopy and ODMR

The depth profile (target depth) of the Fe ion implantation in diamond was modelled with the SRIM package[33] and is shown in Fig. 1(a). Simulations performed for an ion energy of 30 keV, an atomic density of 3.52 g/cm$^3$, and a displacement threshold energy of 45 eV for direction (111)[35], predict that the mean distribution of the depth of implantation is between 5 and 30 nm with the maximum peak at 15.8 nm. The inset to Fig. 1(a) depicts the predicted number of vacancies produced by the Fe ion beam inside the diamond for the ion depth range up to 140 nm, corresponding to the size of the NDs used. For the ND layer (in the form of an air-dried ND suspension) on a flat substrate, the implantation depth corresponds to a local distance from the ND surface. As seen, the entire implantation profile fits within ~50-nm depth, much less than the size of the ND. However, at high-ion fluences used, some damage of the diamond lattice is expected.

To monitor the characteristic structural changes caused by Fe ion irradiation, we recorded fluorescence spectra of the investigated ND samples (Fig. 1(b)). Generally, two competing mechanisms contribute to net light emission, Raman scattering, and photoluminescence. However, for high NV densities and green-light excitation, the specific Raman characteristics are overwhelmed by strong NV fluorescence[31]. To reveal weak Raman contributions on a strong fluorescence background, we performed a baseline correction of the spectra of Fig. 1(b) (fifteen-point spline in Origin software). Fig. 1(c) depicts the resulting corrected spectra with pronounced peaks at wavenumbers of 1426 cm$^{-1}$ (576 nm) and 3129 cm$^{-1}$ (638 nm) that are signatures of zero-phonon lines (ZPL) of NV$^0$, NV$^-$ for nonimplanted and implanted NDs, respectively[36]. Here, the characteristic diamond Raman band corresponding to the active phonon mode F$_{2g}$ at ~1332 cm$^{-1}$ is barely visible in nonimplanted ND and in ND14 due to the overwhelming contribution of highly fluorescent NV centers[31]. The unidentified peak at 2975 cm$^{-1}$ could be due to structural defects, surface chemistry, and/or undefined sample impurities. For the ND15 sample, a broad asymmetric peak is observed between 1200 cm$^{-1}$ and 1700 cm$^{-1}$. This broad structure comes from partially merging peaks due to a combination of disordered and amorphous carbon sp$^2$ (1320 cm$^{-1}$), the 'G-band' of graphitic materials arising from mixing of carbon structures sp$^2$ and sp$^3$ (1570 cm$^{-1}$), and from the contribution of NV$^0$ (~1426 cm$^{-1}$)[37]. At higher irradiation doses, other processes may also affect the surface structure, such as the formation of multiple vacancy defects. Furthermore, with Fe-ions implanted in ND, distinct *blue shifts* of Raman peak positions of NV$^-$ are visible: ~33 cm$^{-1}$ for ND14, and about 110 cm$^{-1}$ shift for ND15 (indicated by arrows in Fig. 1(b) and 1(c)).

To the best of our knowledge, the specific process behind such a large upshift in the wavenumbers is not known. Based on the spectra presented in Figs. 1(b) and 1(c), we attribute the observed large *blue shift* of the Raman spectra caused by the implantation to surface changes in the local lattice environment (e.g., disordered carbon/graphite sp$^2$) that affect the NV$^-$ centers.

In addition to the described changes of the Raman spectra, such as the formation of disordered carbon/graphite sp$^2$, we also observed modifications of the ODMR by ion implantation. Detection of ODMR is based on recording spontaneous fluorescence of excited states of NV centers in a diamond, with a physical mechanism different from that of a Raman scattering where a scattering occurs often without a sizeable population of excited states. Although Raman and fluorescence spectra generally differ[38–40], there are cases, like in NV diamonds, where they overlap and become indiscernible (particularly for high-NV-density samples, see, e.g., Ref. 31). Fig. 2 shows the normalized ODMR spectra recorded in the absence of the magnetic field ($B = 0$) and with applied magnetic fields ($B = 1$ mT and 3 mT). The spectra exhibit a characteristic decrease in fluorescence intensity around the central resonance frequency $D = 2870$ MHz. The decrease is due to repopulation of the m$_S = \pm 1$ sublevels induced by the presence of the MW field. For NDs (Fig. 2(a), at $B = 0$), two resonances caused by strain splitting of about 6 MHz, predicted by the NV spin Hamiltonian[41], are clearly resolved. A small resonance asymmetry is attributed to MW polarization and the frequency dependence of the antenna efficiency[42,43]. The vertical dotted line indicates the ZFS resonance frequency ($D$ = 2870 MHz at $B$ = 0). As can be seen in Fig. 2(a), the ODMR signals with Fe-implanted samples are visibly *red shifted*. The shift increases with increasing implantation dose reaching about 2 MHz for a dose of $10^{15}$ ions/cm$^2$.

The ODMR spectra recorded with $B \neq 0$ are significantly broadened compared to the zero-field spectra, due to random crystallographic orientations of small NDs (Figs. 2(b) and 2(c)). However, the overall width of the broadened spectrum scales with the applied strength of $B$, allowing magnetometric applications[14]. Additionally, there is a small upshift in the central frequency (vertical dotted line) at 2875 MHz (Fig. 2(c), ND14) and an asymmetry in the left and right sides of the spectra that is due to the nonlinear Zeeman effect[44].

To understand the effects of Fe ion implantation on NV centers in NDs, samples were not only analysed using confocal Raman[31] and ODMR spectroscopy. The implanted samples were also analysed using luminescence spectroscopy and spin relaxometry. The corresponding spectra are shown as supporting information. The supporting Fig. S1 shows acquired luminescence spectra, and Figs. S2(a) and S2(b) present the results of measurements of the spin-lattice and spin-spin relaxation times, $T_1$ and $T_2$. These additional studies demonstrate that Fe$^+$ implantation does not impact NV$^-$ luminescence or spin relaxation in a meaningful way.

Implantation introduces damage to the irradiated crystal lattice that can be repaired by thermal annealing[45]. As shown in the inset of Fig. 3(a), the *red shift* of the central frequency of the ODMR signal, caused by irradiation, disappears after annealing. Also, as shown by the photograph in white light in the inset, there is a color change back to the transparent original ND sample. In the Raman spectra (Fig. 3(b)), it is seen that the peaks are restored to their original positions and that the peaks related to the disordered surface bonds of sp$^2$ disappeared. This clearly indicates that, by annealing, the diamond structure is preserved to the original without inclusions of unwanted graphitic structures.

To elucidate the *red shift* observed in the ODMR central resonance frequency, we collected ODMR spectra for ND14 and ND15 as a function of MW and laser powers. From Figs. 4(a) and 4(b), we observe that the ODMR central resonance frequency remains 2870 MHz (ND14) and *red-shifted* to 2865 MHz (ND15) for various applied MW powers between -30 and +10 dBm, at constant laser power of 65 mW. On the other hand, at a constant MW power of -20 dBm, a gradual *red shift* in the ODMR central resonance frequency is observed with changing laser power from 5 to 80 mW as seen in Figs. 4(c) and 4(d). The ODMR spectra and the central resonance frequencies after annealing (Fig. S3) are similar to those for NDs (see Figs. 5(a) and 6(a)). We attribute the *red shift* observed in the NDs implanted with ions (before annealing) to the local thermal effects arising from the heating of the NDs. To provide more insight into the mechanism of that effect, we admixed absorbing powders to the investigated ND samples. Specifically, we prepared samples NDF and NDG where magnetic iron-containing $Fe_2O_3$ and graphite powders created a light-absorbing environment of the investigated NV-NDs and studied their ODMR spectra as functions of varying MW and laser power.

**B. Iron-based and graphitic environment: ODMR**
**B1. Microwave-dependent ODMR signals**

Fig. 5 shows the ODMR signals from NDs by varying the MW power from -30 to +10 dBm, while keeping the laser power constant at 65 mW. Fig. 5(a) displays the resulting ODMR spectrum for NDs as a function of increasing MW power. ODMR signals lead to the observation of a power broadening of the central resonance frequency (2870 MHz) and increase their contrast to saturation by approximately 8% and become power broadened[46,47]. For absorber-enriched NDF and NDG samples, both broadening and contrast increase with increasing MW power (Figs. 5(b) and 5(c)), similar to that observed in Fig. 5(a). Furthermore, for both samples, the strain-split ODMR peaks merge into a single broad fluorescence dip.

Most importantly, the central resonance frequency, 2870 MHz for the sample without admixtures, is *red-shifted* to 2862 MHz for NDF and 2857 MHz for NDG samples due to the presence of iron and graphite admixtures[48].

**B2. Laser power-dependent ODMR signals**

Fig. 6 shows the dependence of the ODMR signal with varying laser power from 5 to 80 mW at -10 dBm MW power. The ODMR signals procured for NDs show no significant changes in the ODMR contrast and/or broadening, and all resonances remain centered at 2870 MHZ with no shift within the applied laser-power range (see Fig. 6(a)). As can be seen from the figure, the applied laser power is sufficiently low to avoid heating of the ND samples with no admixtures. However, for the same laser power that was used to collect ODMR signals for NDs, we observe a significant shift of the central frequency of the ODMR spectra taken with samples with an iron-based and graphitic environment (Figs. 6(b) and 6(c)). In contrast, the spectral shape of the ODMR signals remained unaffected. Here, we used Lorentzian function to determine the central position of the structured resonance. We fitted the two central dips and calculated the average frequency to determine the central resonance frequency '*D*'. For the NDF sample, the ZFS parameter *D* is linearly *red shifted* with increasing laser powder from 2870 MHz at 5 mW to 2859 MHz to 80 mW, and for the NDG sample the shift reached even

2856 MHz. The change from the center position $D$ = 2870 MHz is equivalent to a shift of 11 MHz (NDF) and 14 MHz (NDG) (Fig. 6(d)). The temperature response in NV-ND reveals a thermal shift of ZFS ($D$) at 2870 MHz ($m_s$ = 0 to $m_s$ = ±1) or the zero-phonon line (ZPL) at 637 nm[21]. From Fig. 6(d), this is a direct indication that the frequency shift corresponds to an increase in temperature due to the presence of iron and graphite surroundings, as optically detected using NV⁻ spin resonance. The surrounding environment absorbs laser light, becomes warmer and further heats the NDs, thus increasing the local temperature. For bulk diamonds, the zero-field resonance frequency D changes approximately linearly with temperature around room temperature, with a slope of d$D$/dT= -75 kHz/K[49,21], while the resonance width remains temperature independent[50]. The measured ODMR frequency is temperature dependent, hence, by measuring the ODMR frequency shift, the temperature change can be inferred directly. Taking into account the d$D$/dT reported values, the temperatures are calculated for the frequency shift and plotted as the temperature axis on the right-hand side scale of Fig. 6(d). The calculated temperature shifts appear to be quite strong, exceeding 140 K (NDF) and 180 K (NDG), respectively. The shifts are linear as a function of laser power, as shown in Fig. 6(d), the slope for $Fe_2O_3$ surrounding NDs is ≈ -0.13 MHz/mW, corresponds to the temperature shift ≈ -1.73 K/mW, different from the slope for graphite surrounding ND, which is ≈ -0.17 MHz/mW, and corresponds to ≈ -2.26 K/mW.

## IV. CONCLUSION

In this work, we describe the effects of implantation of Fe ions in ND matrices containing NV⁻ color centers. The implantation becomes responsible for the formation of sp²-structures in NDs resulting in a *blue shift* of Raman spectra. Furthermore, the ODMR signals of NV⁻ are shifted to low frequencies due to local thermal effects caused by a magnetic and non-carbon matter.

We have demonstrated that the ODMR shifts are significantly enhanced by creating optically absorptive iron oxide ($Fe_2O_3$) and graphitic (sp²) environments of the investigated NDs. In this way, we have evidenced extreme temperature changes in NDs that exceed 150 K. We have verified that these changes are not caused by direct heating of non-irradiate NDs by a laser or by microwaves. We attributed these shifts to modification of the local temperature by the absorptive environment of NV-NDs.

The reported ODMR observations shed light on the dynamics of interaction between surfaces of color-center containing nanoparticles with various non-carbon moieties and their temperature dependence. In particular, they reveal the importance of thermal effects associated with various admixtures and/or graphitization in various NV-based experiments, which may enable precision *in situ* thermometry on a nanoscale.

## SUPPLEMENTARY MATERIAL

Fig. S1 shows acquired luminescence spectra, and inset shows amplitudes of the $NV^0$, $NV^-$ peaks and their ratios $NV^-/NV^0$ obtained from the average of ten PL spectra for Fe-ions implanted in ND (ND14 and ND15). Figs. S2(a) and S2(b) present the results of measurements of the spin-lattice and spin-spin relaxation times, $T_1$ and $T_2$ for various Fe-ion implanted doses. Fig. S3 shows the ODMR spectra and the central resonance frequencies after annealing ND14 and ND15. The ODMR signals are collected (a, b) at a laser power of 65 mW at various MW powers and (c, d) with a MW power of -20 dBm at various laser powers.

## ACKNOWLEDGEMENTS

This work was supported by the Foundation for Polish Science TEAM-NET Programme co-financed by the EU under the European Regional Development Fund, grant no. POIR.04.04.00- 00-1644/18.

The authors thank Prof. S. Baran and Dr. Anna M. Majcher-Fitas from the Institute of Physics, and Dr. Anna M. Nowakowska from Faculty of Chemistry, Jagiellonian University, Poland, for supporting this research by providing access to their X-ray diffraction, magnetic measurements, and Raman spectroscopy instrumentation facility.

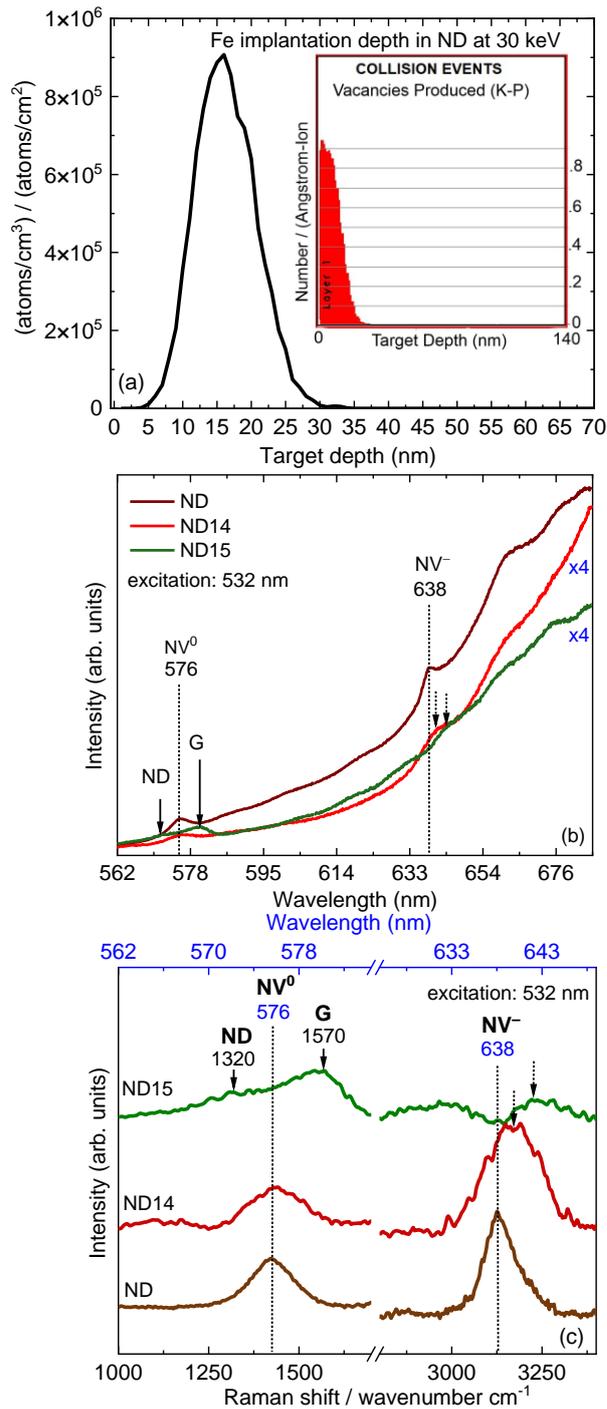

**FIG. 1.** (a) simulated Fe-ion concentration depth profile and damage (inset) at 30 keV. (b) Fluorescence spectra of the Fe implanted 140-nm-sized NDs with: no implantation (brown line), implantation doses of $10^{14}$ ions/cm$^2$ [ND14] (red line) and $10^{15}$ ions/cm$^2$ [ND15] (green line) at 532 nm laser excitation. The marked arrow at ~1320 cm$^{-1}$ is attributed to diamond Raman band in combination to disordered/amorphous carbon, and that at 1570 cm$^{-1}$ to the graphitic material. The dotted lines at 1426 cm$^{-1}$ (576 nm) and 3129 cm$^{-1}$ (638 nm) are signatures of zero-phonon lines (ZPL) of NV$^0$, NV$^-$. The dotted arrows indicate blue shifted NV$^-$ resonance, and (c) shows Raman spectra of NV centers (zoomed-in) after subtraction of a strong fluorescence background.

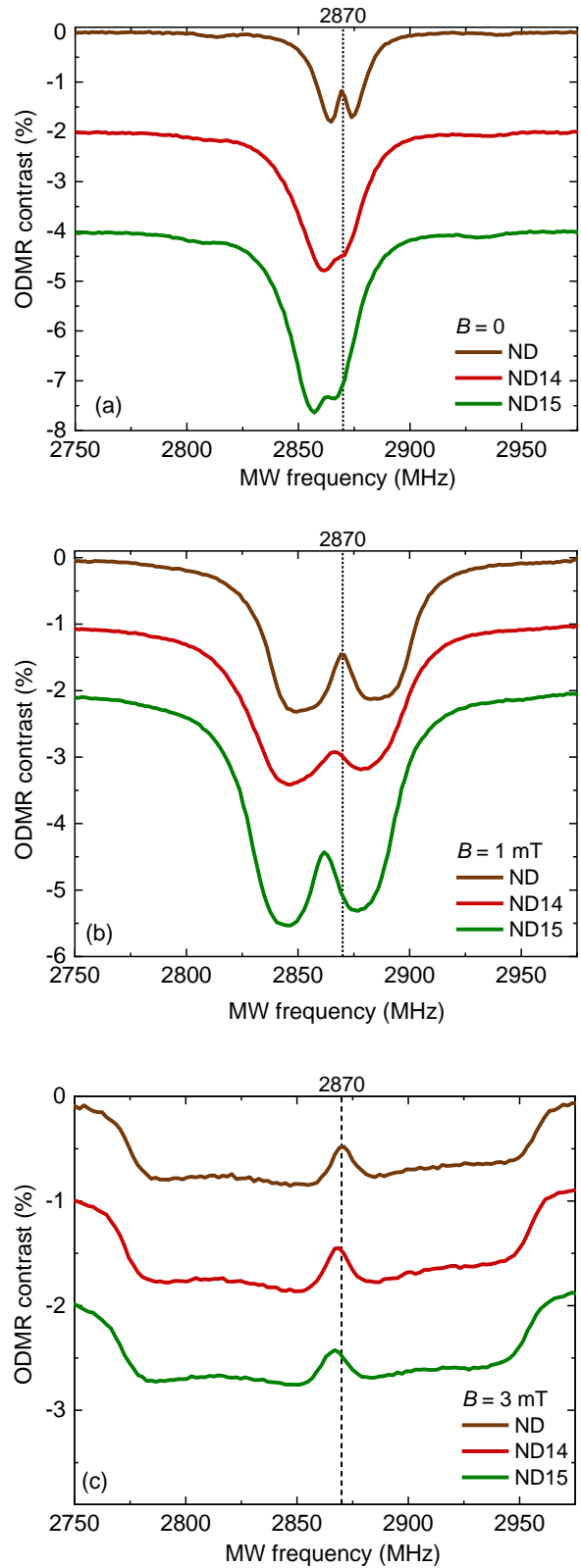

**FIG. 2.** ODMR spectra of ND for various implanted doses and magnetic field intensities: (a) B = 0, (b) B = 1 mT and (c) B = 3 mT.

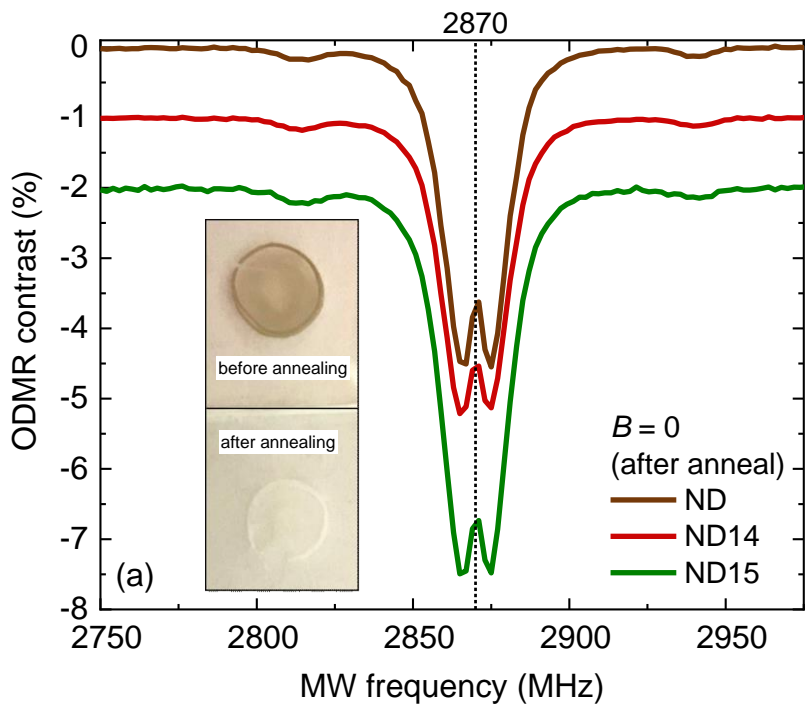

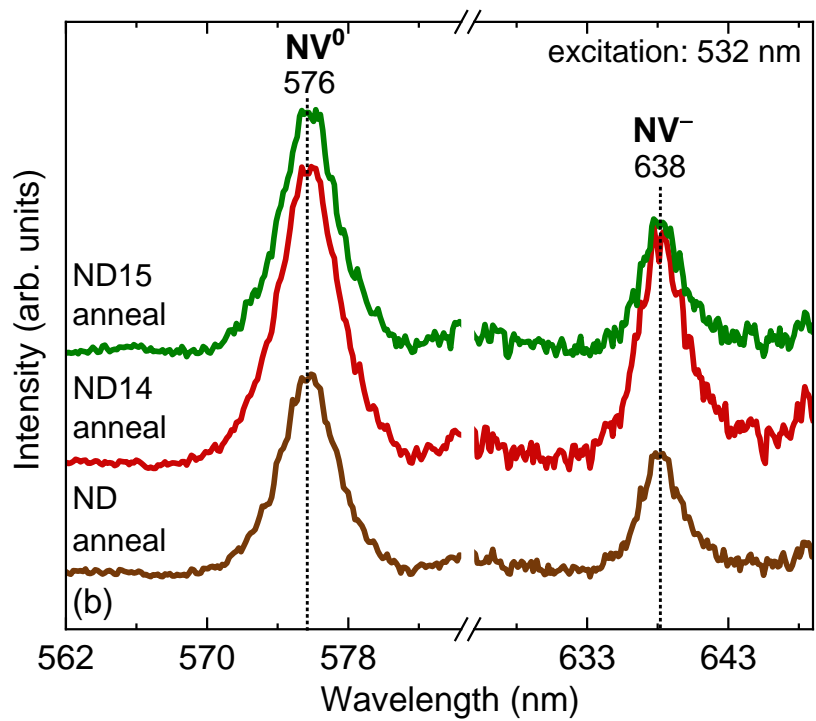

**FIG. 3.** (a) ODMR spectra of ND samples for various implanted doses collected at B = 0 after annealing. The inset shows images of the implanted samples before and after annealing (the samples were drop-dried droplets of 140-nm ND water suspension on a glass surface); (b) Fluorescence spectra of NV centers after implantation and annealing of NDs, with the subtracted strong fluorescence background.

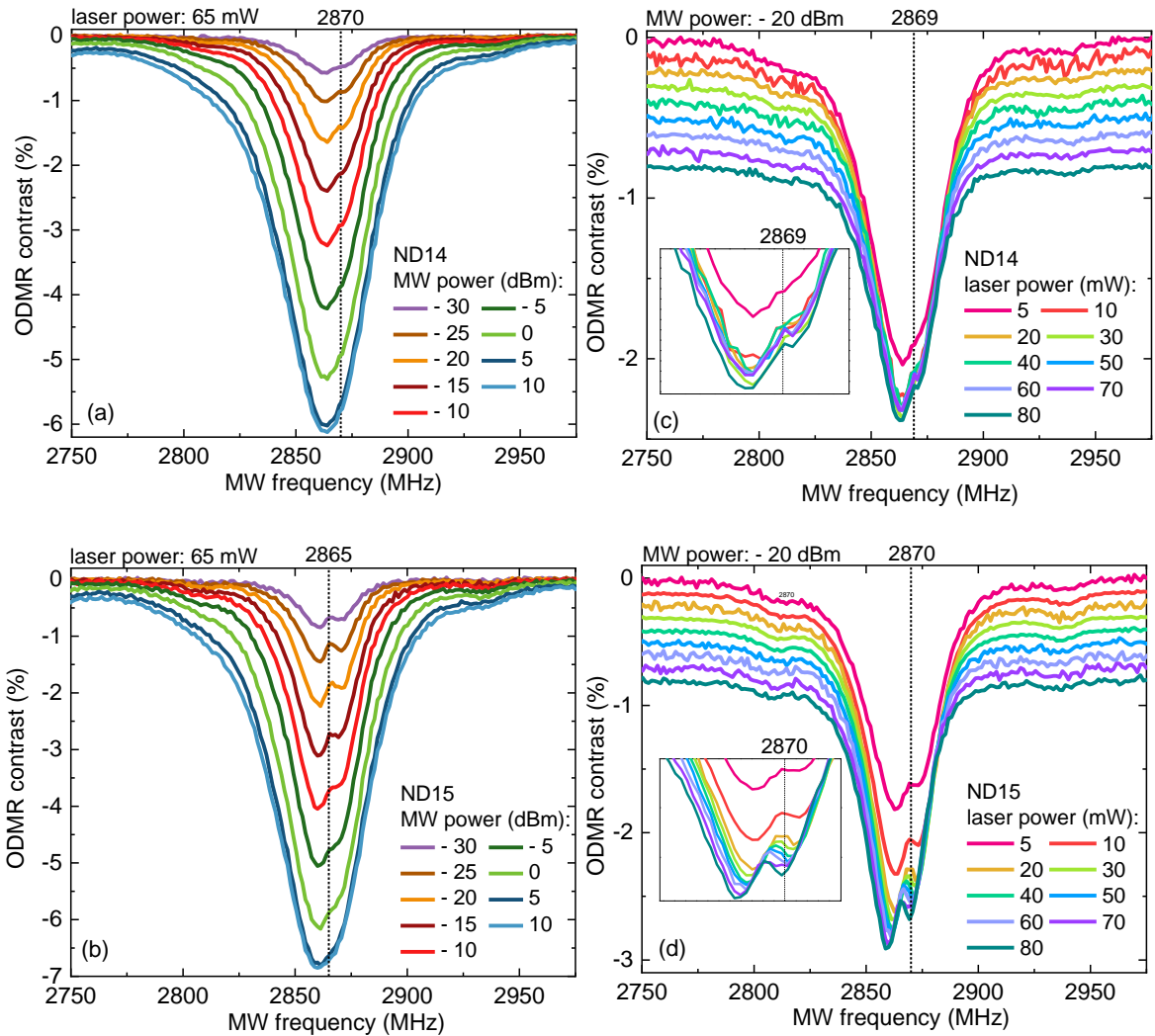

**FIG. 4.** ODMR spectra of ND14 and ND15 before annealing collected (a, b) at a laser power of 65 mW at various MW powers and (c, d) with a MW power of -20 dBm at various laser powers. Insets in (c, d) show the zoomed-in image of the *red shift* observed in the ODMR central resonance frequency.

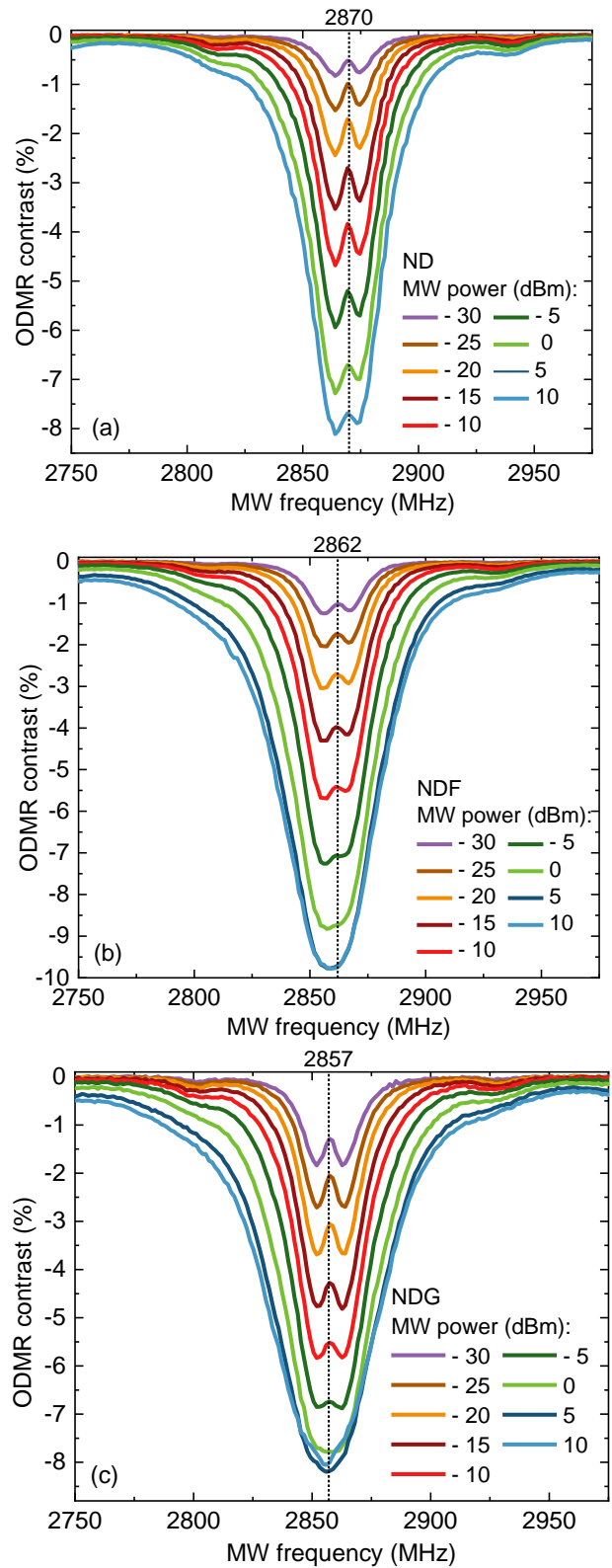

**FIG. 5.** ODMR spectra of (a) ND with admixtures and ND with admixtures of (b) iron ($Fe_2O_3$) and (c) graphite collected at a laser power of 65 mW at various MW powers. Note: In (a) the curves corresponding to 5 dBm and 10 dBm overlap.

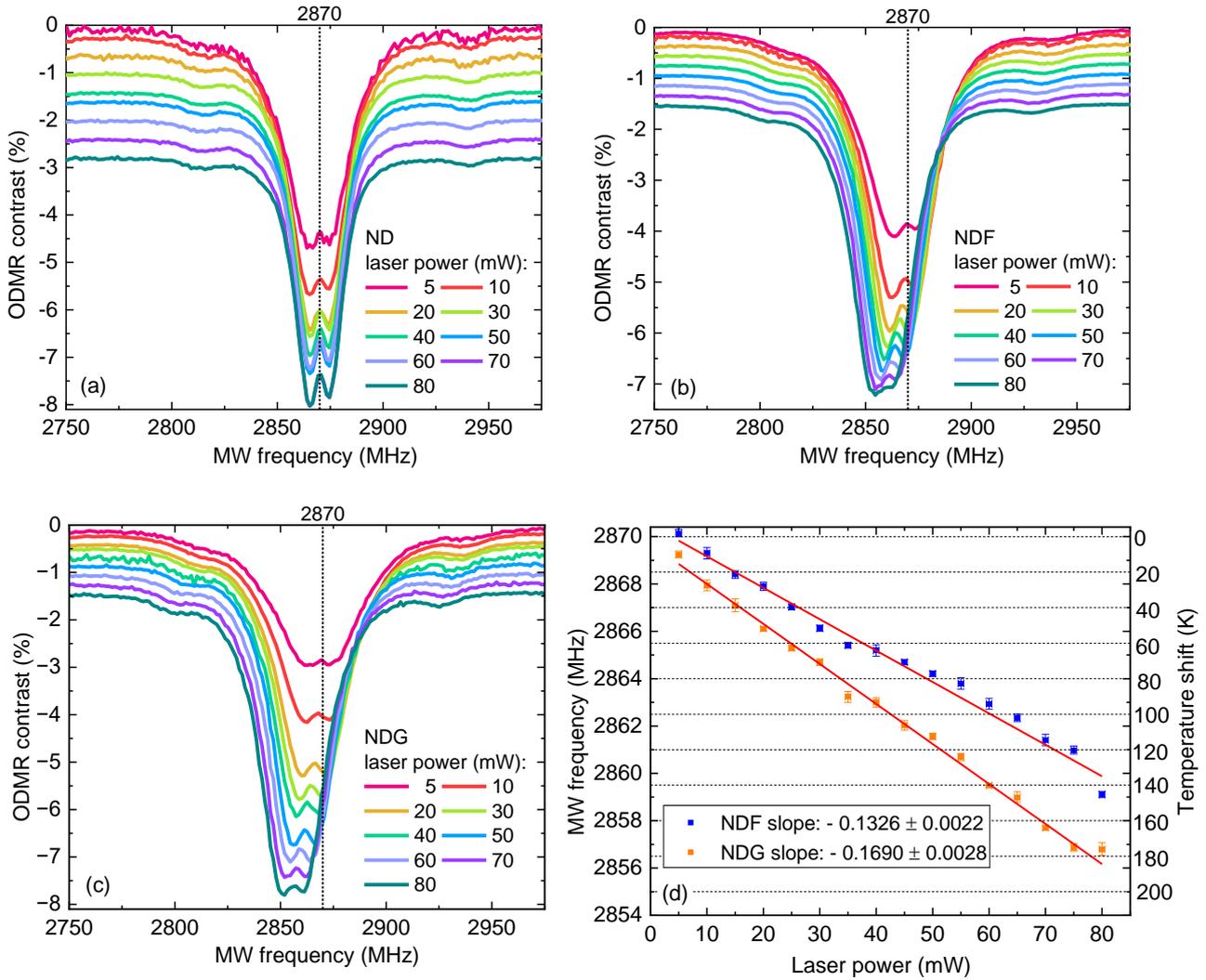

**FIG. 6.** ODMR spectra collected with a MW power of -10 dBm at various laser powers of ND samples (a) without admixtures, and ND with admixtures of (b) iron ($Fe_2O_3$), and (c) graphite. (d) Temperature change due to the local environment around NV-ND calculated from the shift observed in the ODMR spectra (b) and (c). Lorentz fitting was performed on each resonant position in figures (a), (b), and (c), and the values were averaged to calculate the central resonance frequency and errors. The temperature shift was determined using the formula d$D$/d$T$= -75 kHz/K[49,21].